\documentclass[a4paper,11pt]{article}

\usepackage{contribution}

\newcommand{\weblink}[2][]{%
    \ifthenelse{\equal{#1}{}}%
    {\textnormal{\url{#2}}}%
    {\textnormal{\href{#2}{#1}}}%
}

\newcommand{\acknowledgements}[1]{%
  \bigskip\bigskip
  \textsf{\textbf{\Large Acknowledgements}} \\[2ex]
  {#1}
  \bigskip
}

\def\beq{\begin{equation}}
\def\eeq#1{\label{#1}\end{equation}}
\def\eeqn{\end{equation}}

\def\beqa{\begin{eqnarray}}
\def\eeqa#1{\label{#1}\end{eqnarray}}
\def\eeqan{\end{eqnarray}}

\let\bar=\overbar

\def\Dslash{\not{\hbox{\kern-4pt $D$}}}
\def\dslash{\not{\hbox{\kern-2pt $\del$}}}

\def\msb{{\bar{\ssstyle M \kern -1pt S}}}

\newcommand{\contribution}[7][]{%
  \clearpage
  \thispagestyle{plain}
  \ifthenelse{\equal{#1}{}}
  {\hypersetup{pdftitle={#2}}}
  {\hypersetup{pdftitle={#1}}}
  \hypersetup{pdfauthor={{#3} {#4}}}
  {\centering\normalfont\LARGE\bfseries\sffamily #2 \par\nobreak}
  \lhead{}
  \chead{%
    \textit{\footnotesize XIV International Conference on Hadron Spectroscopy
      (\weblink[\textit{hadron2011}]{http://www.hadron2011.de}), 13-17 June 2011, Munich, Germany}%
  }
  \rhead{}
  \bigskip
  \begin{center}
    {#3} {#4}\ifthenelse{\equal{#6}{}}{}{\footnote{\weblink[#6]{mailto:#6}}}
    \ifthenelse{\equal{#7}{}}{}{#7} \\
    \textit{#5}
  \end{center}
  \bigskip
}

\renewcommand{\abstract}[1]{%
  \begin{center}
    \begin{minipage}{0.85\textwidth}
      \begin{footnotesize}
        #1
      \end{footnotesize}
    \end{minipage}
  \end{center}
  \bigskip
}

\begin{document}

%
%
%
%
%
{  

\makeatletter
\@ifundefined{c@affiliation}%
{\newcounter{affiliation}}{}%
\makeatother
\newcommand{\affiliation}[2][]{\setcounter{affiliation}{#2}%
  \ensuremath{{^{\alph{affiliation}}}\text{#1}}}
%

%

\contribution[Partial restoration of chiral symmetry]
{Partial restoration of chiral symmetry \\ and pion in nuclear medium}
{Daisuke}{Jido}  
{\affiliation[Yukawa Institute for Theoretical Physics, Kyoto University, Kyoto 606-8502, JAPAN]{1} \\
 \affiliation[Department of Physics, Graduate School of Science, Kyoto University,
Kyoto, 606-8502, JAPAN]{2} \\
}
{jido@yukawa.kyoto-u.ac.jp} 
{\!\!$^,\affiliation{1}$, and Soichiro Goda\affiliation{2}}
%
%

\abstract{%
We discuss partial restoration of chiral symmetry in nuclear medium, which is realized as 
an effective reduction of the quark condensate in nuclear medium. We derive the in-medium 
Weinberg-Tomozawa relation. We also give a brief calculation of the in-medium quark condensate based on chiral perturbation theory. 
We conclude that the density dependence of the quark condensate beyond the linear 
density comes from the vertex correction of the chiral field insertion and nucleon correlations. 
}
%

\section{Introduction}

One of the goals of the contemporary nuclear physics is to figure out the QCD vacuum structure at finite density and/or temperature. Especially, the fate of dynamical breaking of chiral symmetry in finite baryonic density is phenomenologically interesting, because we could obtain experimental evidence of partial restoration of chiral symmetry in the nuclear medium by investigating in-medium properties of meson in nuclei. 
Recently, precise measurements of the spectra of deeply bound 
pionic atoms were performed~\cite{deeppiexp,Suzuki:2002ae}, and with these data
the pion optical potential parameters were determination in detail.  
Especially, the repulsive enhancement of 
the isovector $\pi^{-}$-nucleus interaction was accurately 
extracted as $b_{1}^{\rm free}/b_{1} = 0.78 \pm 0.05$
at around $\rho\sim 0.6 \rho$~\cite{Suzuki:2002ae}. 
The $b_{1}$ repulsive enhancement was also seen in low-energy
pion-nucleus scatterings~\cite{Friedman:2004jh}.
With help of the theoretical discussion~\cite{Kolomeitsev:2002gc,Jido:2008bk},
the $b_{1}$ parameter is converted to the in-medium pion decay constant $F_{t}$,
and we concluded that the experimental finding of the $b_{1}$ enhancement 
is to be a signal of the reduction of the pion decay constant in nuclear matter. 
Further, the exact relation between the in-medium pion decay constant and
quark condensate was also found at the linear density approximation 
in Ref.~\cite{Jido:2008bk}. Now we have arrived at the qualitative confirmation 
of the partial restoration of chiral symmetry in nuclear medium and should go a
step further to make the argument more quantitative~\cite{Kaiser:2007nv,Ikeno:2011mv}. 
In this paper we briefly discuss the sum rule for the in-medium quark condensate and
the in-medium Weinber-Tomozawa relation. We also show a calculation of the 
in-medium quark condensate based on chiral perturbation theory beyond the linear density.

\section{Sum rule for the in-medium quark condensate}

To connect the phenomenological consequences extracted from experimental
observation to the quark condensate, we need theoretical consideration which 
makes bridge between hadronic description and quark language. 
In Ref.~\cite{Jido:2008bk}, an exact sum rule for a symmetric nuclear matter 
was derived in the chiral limit:
\begin{equation}
    \sum_{\alpha} {\rm Re} \left[ F_{t,\alpha} G_{\alpha}^{*1/2} \right] 
    = - \langle \bar qq \rangle^{*} \label{eq:SR}
\end{equation}
where the summation is taken over all of pionic zero modes in nuclear matter which
have the same quantum number with the pion in nuclear matter and whose energy 
is zero measured from the nuclear matter ground state, and the matrix elements 
of $F_{t,\alpha}$ and $G^{1/2}_{\alpha}$ are given in the nuclear matter rest frame by
\begin{eqnarray}
   \langle \Omega | A_{0}^{a}| \Omega_{\alpha}^{b} (k) \rangle &=& i \delta^{ab} \omega_{\alpha} F_{t,\alpha}  \label{eq:ft} \\
   \langle \Omega^{b}_{\alpha} | \phi_{5}^{a} | \Omega \rangle  &=& \delta^{ab} G_{\alpha}^{*1/2}
\end{eqnarray}
with the axial current $A_{\mu}^{a}$, the pseudoscalar density $\phi_{5}^{a}$, the ground state of the symmetric nuclear matter $ | \Omega \rangle$ and the pionic zero mode state $| \Omega_{\alpha} \rangle$. 

One of the most important consequence of this sum rule is that to obtain the in-medium
chiral condensate one has to sum up all of the pionic zero modes. This means that 
one should not have to separate out the in-medium pion properties from complicated
dynamics of pion and nuclear matter. 
This sum rule is valid for all densities and derived by current algebra as a low energy theorem. 
Instead, we need 
description of dynamics of in-medium pion and nuclear matter for actual calculation
of matrix elements. This sum rule is also available for experimental confirmation of 
partial restoration of chiral symmetry, once the matrix elements are  
extracted from experimental observation. 

In the linear density approximation, the sum rule can be simplified to 
$
    F_{t} G_{\pi}^{*1/2} = - \langle \bar qq \rangle^{*} 
$.
By taking its ratio to the in-vacuum Glashow-Weinberg relation $F_{\pi} G_{\pi}^{1/2} = -
\langle \bar qq \rangle$~\cite{Glashow:1967rx}, we obtain the scaling law
\begin{equation}
    \frac{F_{t}}{F_{\pi}} Z_{\pi}^{1/2} = \frac{\langle \bar qq \rangle^{*}}{\langle \bar qq \rangle} \ .
\end{equation}
where $Z^{1/2}_{\pi}$ is the in-medium wavefunction renormalization, which can be 
extracted at the linear density from the $\pi N$ scattering data.
The in-medium reduction of the pion decay constant was obtained in the pionic atom. 

\section{In-medium Weinberg-Tomozawa relation}

As discussed in Ref.~\cite{Jido:2008bk}, the in-medium Weinberg-Tomozawa
relation can be derived by considering the chiral-limit correlation function of 
the axial current $A_{\mu}^{a}$ in the asymmetric nuclear matter:
$
  \Pi_{\nu}^{ab}(q) = \int d^{4} x e^{iq\cdot x} \partial^{\mu} 
  \langle \Omega^{\prime} | T [A_{\mu}^{a}(x) A_{\nu}^{b}(0)] | \Omega^{\prime} \rangle,
$
where $|\Omega^{\prime}\rangle$ is the ground state of the asymmetric 
nuclear matter normalized as 
$\langle \Omega^{\prime} | \Omega^{\prime} \rangle =1$ and specified by 
the isoscalar density  $\rho=\rho_{p} + \rho_{n}$ and the isovector density 
$\delta \rho = \rho_{p} - \rho_{n}$. In the soft limit, using the axial current 
conservation $\partial \cdot A =0$, we obtain
\begin{equation}
   \Pi_{\nu=0}^{ab}(0) = \int d^{3}x [A_{0}^{a}(x), A_{0}^{b}(0)] 
   = i \epsilon^{ab3} \langle \Omega^{\prime} | V_{0}^{3} | \Omega^{\prime} \rangle,
   \label{eq:AAcorr}
\end{equation}
where we have used the commutation relation $[Q_{5}^{a}, A_{\nu}^{b}] = i \epsilon^{abc} V_{\nu}^{c}$. The matrix element in the right hand side of Eq.~(\ref{eq:AAcorr}) implies a spacial average of the isospin density in the nuclear matter and counts the $z$ component of the isospin of the nuclear matter state. For the ground state of nuclear matter, at the linear $\delta \rho$, the matrix element is written as
\begin{equation}
    \Pi^{ab}_{0}(0) \simeq i \epsilon^{ab3} \frac{1}{2} \delta \rho .
\end{equation}
On the other hand, inserting the hadronic complete set into the correlation function, we obtain the hadronic description of the correlation function in the soft limit
where we take $\vec q \to 0$ first as
\begin{equation}
   \Pi_{0}^{ab}(0) = \lim_{\omega \rightarrow 0} i \omega (\omega F^{a\alpha}_{t})
   \frac{1}{\omega^{2} \delta^{\alpha\beta} - \Sigma^{\alpha\beta}} (\omega F^{\beta b}_{t}) , 
\end{equation}
where $\Sigma^{\alpha\beta}$ is the self-energy of the zero mode and $F_{t}^{ab}$ is the matrix element of the axial current for the ground state
$| \Omega^{\prime} \rangle $ and the pionic zero modes in the asymmetric nuclear 
matter $|\Omega^{\prime}_{\alpha} \rangle$ given by
$
    \langle \Omega^{\prime} | A_{0}^{a} | \Omega^{\prime}_{\alpha} \rangle 
    = i \omega F_{t}^{a\alpha} 
$.
Thus we obtain a sum rule
\begin{equation}
  \sum_{\alpha,\beta}  \lim_{\omega\to 0} \omega F_{t}^{a \alpha} 
  \left( \delta^{\alpha\beta} - \frac{1}{2\omega} 
  \frac{\partial \Sigma^{\alpha\beta}}{\partial \omega} \right)^{-1}  F_{t}^{\beta b} = \epsilon^{ab3} \frac{1}{2} \delta \rho .
\end{equation}
The explicit expression of the $F_{t}^{a\alpha}$ and $\Sigma^{\alpha\beta}$ depends
on the description of the complete set, but the sum is not dependent on the description. 
Now we expand $F_{t}^{\alpha\beta}$ and $\Sigma^{\alpha\beta}$ 
in terms of the isovector density $\delta\rho$, and find  
\begin{equation}
   \Sigma^{\alpha\beta} =  \epsilon^{\alpha\beta 3} \frac{\omega}{ F_{t}^{2}} \delta \rho,
\end{equation}
where $F_{t}$ is the decay constant in the symmetric nuclear matter given 
in Eq.~(\ref{eq:ft}). This is just a consequence of the isospin symmetry 

\section{A model calculation of the quark condensate}

The sum rule (\ref{eq:SR}) was derived by considering the  correlation
function of the axial current and pseudoscalar density in symmetric nuclear matter 
at the chiral limit:
$
    \Pi_{5}^{ab}(q) =  \int d^{4}x e^{iq\cdot x} \partial^{\mu} 
    \langle \Omega | T[A_{\mu}^{a}(x) \phi_{5}^{b}(0) ] | \Omega \rangle.
$
Taking the soft limit and using the Ward-Takahashi identity, we obtain
  \begin{equation}
    \lim_{q \to 0} \int d^{4}x e^{iq\cdot x} \partial^{\mu} 
    \langle \Omega | T[A_{\mu}^{a}(x) \phi_{5}^{b}(0) ] | \Omega \rangle
    = - i \delta^{ab} \langle \bar qq \rangle^{*}
  \end{equation}
In this section, we directly calculate the left hand side based on 
in-medium chiral perturbation theory.
Here we take, for example, the formulation developed in Refs.~\cite{Oller:2001sn},
in which the generating functional of the SU(2) chiral Lagrangian in non-interacting
nucleon gas environment is calculated in expansions of Fermi-see insertion and 
chiral order counting. The Fermi momentum $k_{f}$ is also regarded as a small 
expansion parameter as well as the pion momentum and mass in the in-vacuum 
chiral perturbation theory. 

  \begin{figure}[t]
    \begin{tabular}{cc}
      \begin{minipage}{0.5\hsize}
        \begin{center}
        \includegraphics[width=6.5cm]{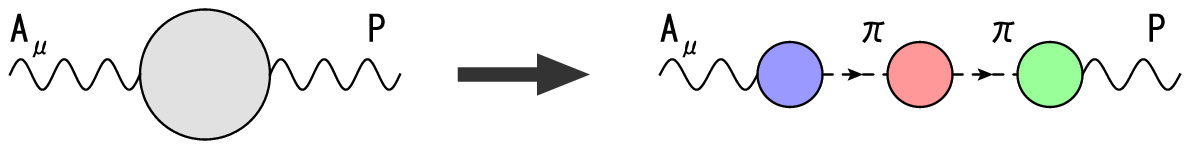}
        \caption{Separation of $ \langle A^a_{\mu}(q) P^b(0) \rangle^* $. }
        \label{AP}
        \end{center}
      \end{minipage}
      \begin{minipage}{0.5\hsize}
        \begin{center}
        \includegraphics[width=6.5cm]{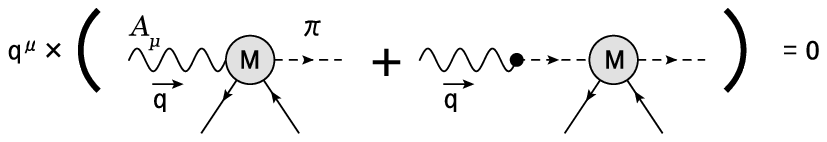}
        \caption{Conservation of  axial current}
        \label{M}
        \end{center}
      \end{minipage}
    \end{tabular}
  \end{figure}

We calculate the in-medium Green function $ \langle A^a_{\mu}(q) P^b(0) \rangle^* $
in the chiral limit~\cite{Goda}. First of all, it is possible to separate the Green function 
into the three parts, the medium corrections of the decay constant, pion wavefunction
and the pseudo-scalar coupling (Fig.~\ref{AP}), which are connected with the in-vacuum
pion propagators. Among three parts, there is cancellation between the corrections
for the decay constant and wavefunction at the soft limit $q_{\mu} \to 0$
thanks to the axial vector current conservation in the chiral limit, 
$ \partial^{\mu} A_{\mu}^a (x) = 0 $. This is a generalized Goldberger-Treiman 
relation (Fig.~\ref{M}). Therefore,  only the density dependence of the 
pseudo-scalar coupling contributes to
the in-medium quark condensate.

\begin{figure}[b]
\begin{center}
       \includegraphics[width=13cm]{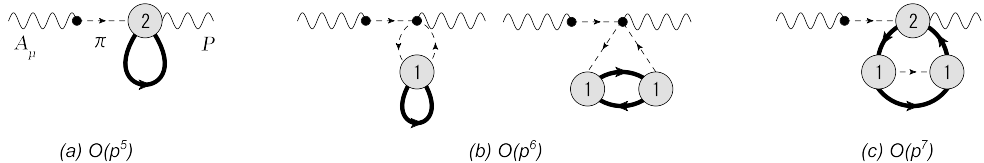}
\end{center}
        \caption{Diagrams for the medium corrections of the quark condensate in 
        the chiral limit. $O(p^{n})$ means the order of the chiral expansion.  }
        \label{inQQ}
\end{figure}

The medium correction for the quark condensate starts from $ O(p^4 )$
of the chiral expansion, but the contribution of this order vanishes at the chiral 
limit due to the cancelation mentioned above. The leading order correction 
is given by the diagram shown in Fig.~\ref{inQQ}(a) with the $O(p^{2})$ $\pi P NN$
vertex. 
This contribution reproduces the well-known result of the linear density approximation~\cite{Drukarev:1991fs}:
\begin{equation}
    \langle \bar{u}u +\bar{d} d  \rangle^* = 
    \langle \bar{u}u + \bar{d}d \rangle_0 (1+ \frac{8 c_1 \rho}{f_{\pi}^2}) ,
\end{equation}
where $\langle \bar{u}u + \bar{d}d \rangle_0$ is the quark condensate in vacuum,
$\rho$ is the nuclear matter density and
$ c_1 $ is one of the low energy constants (LEC) in the $\pi N$ chiral Lagrangian 
at $ O(p^2)$, which is related to the $\pi N$ sigma term. 
With $C_{1}=-0.58$ GeV$^{-1}$,  $f_{\pi}=92.4$ MeV and the normal nuclear density 
$ \rho_0 = 0.17$ fm$^{-3}$, we have
$ \langle \bar{u}u +\bar{d} d  \rangle^* = \langle \bar{u}u + \bar{d}d \rangle_0 (1- 0.35 \rho/\rho_{0})$. 
This suggests 30 percent reduction of the quark condensate at the nuclear density.

Further density corrections beyond the linear density can be calculated with
the diagrams shown in Fig.~\ref{inQQ}(b) and(c). 
They are counted as $O(p^{6})$ and $O(p^{7})$ 
of the chiral expansion and give $k_{f}^{4}$ and higher power contributions. 
In this way, one can perform the systematic expansion in terms of the chiral 
counting with this formulation. Since, as discussed above, the finite density contribution 
to the quark condensate comes only from the medium corrections of
the pseudo-scalar coupling, further higher order corrections come from 
diagrams shown in Fig.~\ref{Higher}. 
This means that the density corrections of the quark condensate 
are given by multi-pion exchanges between nucleons and multi-nucleon correlations
in the nuclear matter. Therefore, in order to perform the realistic calculation 
of the in-medium quark condensate, one has to first formulate realistic nuclear 
matter. Indeed, the in-medium chiral perturbation theory has the systematic 
scheme for counting the order of the chiral expansion and this is good for
theoretical analyses.  Nevertheless,
only the pion-nucleon dynamics described by the chiral perturbation theory
may not describe realistic nuclear matter, and more phenomenological
descriptions are necessary to obtain realistic nuclear matter having 
the saturation properties. Thus, we may have to go beyond the chiral 
counting scheme in order to calculate the quark condensate in nuclear matter. 

  \begin{figure}[t]
  \begin{center}
          \includegraphics[width=13cm]{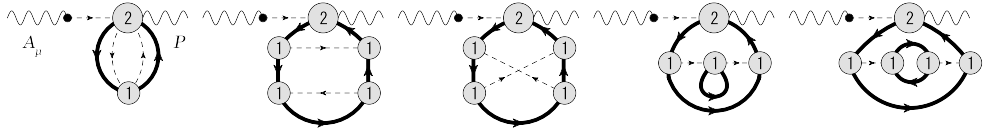}
  \end{center}
        \caption{Further higher order contributions. These diagrams represent the two-pion exchange and nucleon-nucleon correlations. }
        \label{Higher}
  \end{figure}

\section{Conclusion}

The deeply bound pionic atoms are the most successful systems to investigate the 
in-medium pion properties, because we have certainly the bound states with 
so narrow widths that we can perform detailed spectroscopy. From the 
observed spectra we can extract in-medium pion properties. With these 
quantities  we have concluded that partial restoration of chiral symmetry 
takes place in nucleus with help of theoretical arguments, which complete 
the story from the observation to QCD. Now we are going to the next stage 
to determine the in-medium quark condensate quantitatively. 
We have also performed brief calculation of the in-medium quark condensate
based on chiral perturbation theory. This calculation shows that 
to obtain the realistic quark condensate in a nuclear medium 
one needs realistic description of nuclear matter. 

%


\acknowledgements{%
This work was partially supported by the Grant-in-Aid for Scientific Research from 
JSPS (Nos.\ 22740161, 22105507).
This work was done in part under the Yukawa International Program for Quark-hadron Sciences (YIPQS). The Feynman diagrams in this paper were drawn using jaxodraw~\cite{Binosi:2003yf}.}


%

}  


\end{document}